\documentclass[conference]{IEEEtran}
\IEEEoverridecommandlockouts

\usepackage{cite}
\usepackage{amsmath,amssymb,amsfonts}
\usepackage{algorithmic}
\usepackage{graphicx}
\usepackage{subfig}
\usepackage{textcomp}
\usepackage{xcolor}
\usepackage{tikz}
\def\BibTeX{{\rm B\kern-.05em{\sc i\kern-.025em b}\kern-.08em
    T\kern-.1667em\lower.7ex\hbox{E}\kern-.125emX}}

\usepackage{booktabs}

\newcommand{\norm}[1]{\left\lVert#1\right\rVert}

\begin{document}

\title{FlowMAC: Conditional Flow Matching for Audio Coding at Low Bit Rates\\

}

\author{\IEEEauthorblockN{Nicola Pia\textsuperscript{\textdagger}$^1$\thanks{\textsuperscript{\textdagger}Equal contribution to this work}}
\and
\IEEEauthorblockN{Martin Strauss\textsuperscript{\textdagger}$^2$}
\and
\IEEEauthorblockN{Markus Multrus$^1$}
\and	
\IEEEauthorblockN{Bernd Edler$^2$}
\and
Affiliations: \IEEEauthorblockA{Fraunhofer IIS$^1$, Erlangen, Germany. International Audio Laboratories Erlangen$^{*,2}$\thanks{$^*$A joint institution of the Friedrich-Alexander-Universität at Erlangen-Nürnberg (FAU) and Fraunhofer IIS.}, Erlangen, Germany.}
\and
\{nicola.pia,markus.multrus\}@iis.fraunhofer.de, \{martin.strauss,bernd.edler\}@audiolabs-erlangen.de
} 

\maketitle

\begin{abstract}
This paper introduces FlowMAC, a novel neural audio codec for high-quality general audio compression at low bit rates based on conditional flow matching (CFM).
FlowMAC jointly learns a mel spectrogram encoder, quantizer and decoder.
At inference time the decoder integrates a continuous normalizing flow via an ODE solver to generate a high-quality mel spectrogram.
This is the first time that a CFM-based approach is applied to general audio coding, enabling a scalable, simple and memory efficient training.
Our subjective evaluations show that FlowMAC at 3\,kbps achieves similar quality as state-of-the-art GAN-based and DDPM-based neural audio codecs at double the bit rate. 
Moreover, FlowMAC offers a tunable inference pipeline, which permits to trade off complexity and quality.
This enables real-time coding on CPU, while maintaining high perceptual quality.

\end{abstract}

\begin{IEEEkeywords}
neural audio coding, conditional flow matching, low bit rate coding
\end{IEEEkeywords}

\section{Introduction}

In the modern digital world, audio codecs are used on a day-to-day basis, so every technological advancement can have a large impact.
In recent years, deep neural networks (DNNs) revolutionized the field of audio compression.
Early approaches~\cite{firstnn_coding,knet_coding,zhen2020efficientscalableneuralresidual} control the compression via entropy-based losses and ensure good quality via reconstruction losses. 
With the advent of deep generative models the quality of neural codecs at bit rates lower than 12\,kbps greatly improved. 

While for speech coding many different approaches were proven to be successful~\cite{wavenet_coding,lpcnet_coding,ssmgan,cascade_coding}, the general audio codec SoundStream~\cite{soundstream} established a new paradigm of training a residual VQ-VAE~\cite{vq_vae} via an additional GAN loss end-to-end (e2e).
For this, a DNN-encoder extracts a learned latent, a residual VQ generates the bit stream, and a DNN-decoder synthesizes the audio.
All the modules are jointly learned via a combination of multiple spectral reconstruction, VQ-VAE codebook and commitment and adversarial losses.

Various improvements on the design of SoundStream were proposed afterwards.
EnCodec~\cite{defossez2023encodec} used recurrent networks and an improved compression capability via entropy coding based on language models in the quantizer.
The Descript-Audio-Codec (DAC)~\cite{dac} achieved high quality extending on the model size, using innovative audio-specific activations~\cite{snake}, and scaling up the discriminator architecture.

The e2e VQ-GAN approach offers a great flexibility in the design and complexity of the codec~\cite{tfnet,pia22_interspeech,funcodec}.
However, it often entails a complicated and unstable training pipeline, which sometimes fails to meet quality expectations for challenging signal types, particularly at bit rates lower than 6\,kbps.

Denoising Diffusion Probabilistic Models (DDPMs) were proposed recently for speech~\cite{ladiffcodec} and general audio~\cite{NEURIPS2023_MDB,liu2024semanticodecultralowbitrate}.
While~\cite{liu2024semanticodecultralowbitrate} targets semantic coding at ultra low bit rates, MultiBandDiffusion (MBD)~\cite{NEURIPS2023_MDB} is a decoder model that enables high-quality synthesis of the EnCodec latent at 1.5, 3 and 6\,kbps for general audio.
This model uses a time-domain subband-based decoding scheme and achieves state-of-the-art quality for music.
The high complexity of this model makes it hard to use in embedded devices and its dependency on a pre-trained bit stream might limit its compression capabilities.

VQ-GANs entail a highly involved training pipeline and the existing DDPMs are computationally heavy models.
This demonstrates the need for a solution that is easy to train, while offering high quality performance at acceptable complexity.

Recently, a new paradigm to train continuous normalizing flows (CNFs) called conditional flow matching (CFM) emerged~\cite{lipman2023flow} and demonstrated state-of-the-art quality for both image~\cite{esser2024scalingrectifiedflowtransformers} and audio generation~\cite{matcha, p_flow, le2023voicebox}.
This approach offers a simple training pipeline at much lower inference and training costs compared to DDPMs. 

In this work, we present the \textbf{Flow} \textbf{M}atching \textbf{A}udio \textbf{C}odec (\textbf{FlowMAC}), a new audio compression model for low bit rate coding of general audio at $24$\,kHz audio based on CFM.
Our proposed approach learns a mel spectrogram encoder, residual VQ, and decoder via a combination of a simple reconstruction loss and the CFM objective.
The CFM-based decoder generates realistic mel spectrograms from the discrete latent, which is then converted to waveform domain via an efficient version of BigVGAN~\cite{lee2023bigvgan}.
The model design is simple and the training pipeline is stable and efficient.

Our contributions can be summarized as follows: 
\begin{itemize}
	\item We introduce FlowMAC, a CFM-based mel spectrogram codec offering a simple and efficient training pipeline.
	\item Our listening test results demonstrate that FlowMAC achieves state-of-the-art quality at 3\,kbps matching GAN-based and DDPM-based solutions at double the bit rate.
	\item We propose an efficient version of FlowMAC capable of coding at high quality and faster than real time on a CPU.
\end{itemize}

\begin{figure*}[htb!]
	\centering
	\makebox[0.3\textwidth][c]{
		\input{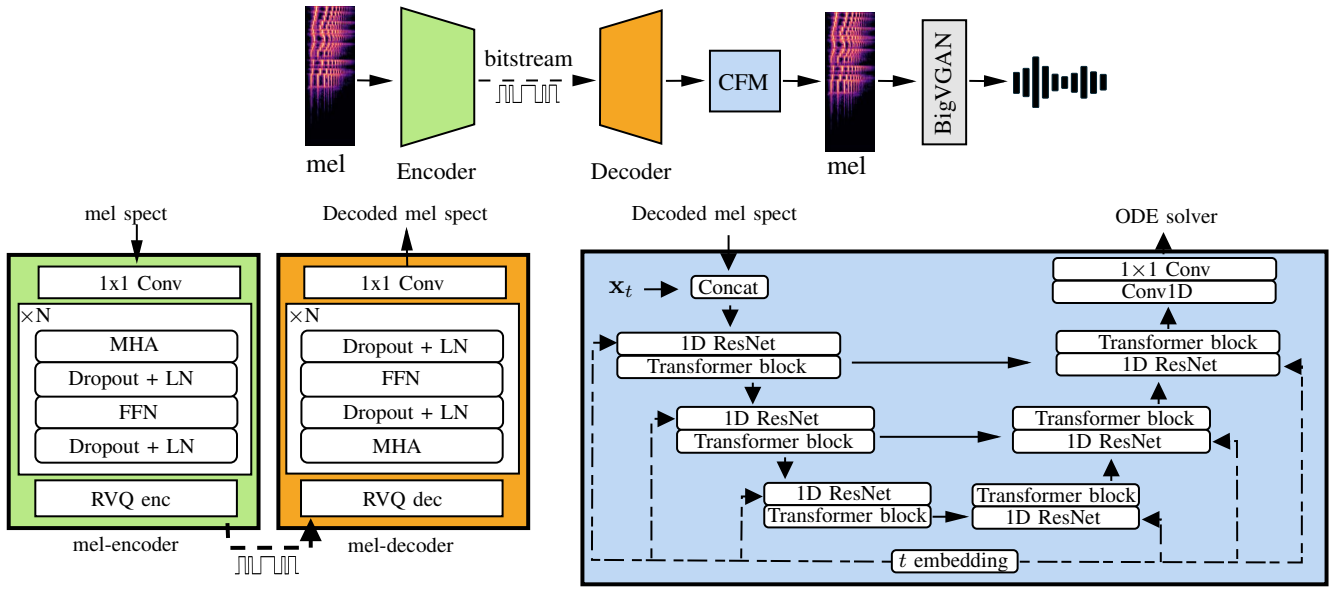}}
	\caption{FlowMAC architecture. The top illustrates the high level pipeline. The bottom left shows the structure of the mel spectrogram encoder and decoder. The bottom right denotes the details on the CFM module.}
	\label{fig:arch}
\end{figure*}

\section{Flow Matching fundamentals}
\label{sec:fm}
For neural audio coding, we learn an encoder-decoder architecture that compresses input mel spectrograms into a quantized bit stream.
We then use the information from this bit stream to condition a CFM-based mel spectrogram decoder for high-quality mel spectrogram generation.
To this end, we consider the distribution $q$ of mel spectrograms of the input audio signals and we learn a time-dependent vector field $\mathbf{u}_t$, whose flow transforms a Gaussian prior $p_0$ into $q$.

Flow matching~\cite{lipman2023flow} describes a method to fit a time-dependent probability density path $p_t:[0,1] \times \mathbb{R}^d \rightarrow \mathbb{R}^{\ge 0}$ between a simple sampling distribution $p_0(\mathbf{x})$ and the target data distribution $q(\mathbf{x})$, where $t \in [0,1]$ and $\mathbf{x} \in \mathbb{R}^d$.
More precisely it defines a framework to train a CNF $\phi_t$ via learning its associated vector field $\mathbf{u}_t$ directly.

Following Section 4.1 in~\cite{lipman2023flow} we define 
\begin{equation}
p_t(\mathbf{x}|\mathbf{x}_1) = \mathcal{N}\left(\mathbf{x}; \mu_t(\mathbf{x}_1), \sigma_t(\mathbf{x}_1)^2\mathbf{I}\right),
\end{equation}

\noindent where $\mathbf{x}_1\sim q(\mathbf{x}_1)$ sampled from the train set, $\mu_t(\mathbf{x}_1) = t \mathbf{x}_1$, and $\sigma_t(\mathbf{x}_1) = 1 - (1 - \sigma_\text{min})t$ with $\sigma_\text{min} \ll 1$.
This defines a Gaussian path where $p_0$ is the standard Gaussian and $p_1$ is a Gaussian centered at $\mathbf{x}_1$ with small variance.
Theorem 3 in \cite{lipman2023flow} shows that this probability path is generated by the Optimal Transport Conditional Vector Field

\begin{equation}
\mathbf{u}_t(\mathbf{x}|\mathbf{x}_1) = \frac{\mathbf{x}_1 - (1 - \sigma_\text{min})\mathbf{x}}{1 - (1 - \sigma_\text{min})t}.
\end{equation}
This yields the conditional flow matching objective
\begin{align*}
	\mathcal{L}_{\textup{CFM}}(\theta) &= \mathbb{E}_{t,q(\mathbf{x}_1),p_t(\mathbf{x}|\mathbf{x}_1)} \norm{\mathbf{v}_t(\mathbf{x};\theta) - \mathbf{u}_t(\mathbf{x}|\mathbf{x}_1)}^2\\
	& = \mathbb{E}_{t,q(\mathbf{x}_1),p_0(\mathbf{x}_0)} \norm{\mathbf{v}_t(\mathbf{x};\theta) - \left(\mathbf{x}_1 - (1 - \sigma_\text{min})\mathbf{x}_0\right)}^2
\end{align*}

\noindent where $\mathbf{v}_t(\mathbf{x}, \theta)$ denotes a DNN parametrized by $\theta$, the time step $t \sim \mathbb{U}[0,1]$ is sampled from a uniform distribution.

For our system the neural network $\mathbf{v}_t(\mathbf{x};\theta)$ is additionally conditioned on the decoded bit stream $c$ obtained from a learned mel spectrogram compression network.
During inference, $\mathbf{v}_t$ takes $c$ as input and a Gaussian noise sample $\mathbf{x}_0$ and outputs the derivatives of the corresponding CNF.
This flow is then integrated using an ODE solver, e.g. the Euler method.

\section{Proposed Architecture}

The architecture of FlowMAC is illustrated in Figure~\ref{fig:arch}.

\subsection{Mel Encoder-Decoder}
\label{subsec:enc-dec}
The $128$ mel spectrogram bands are calculated on the input $24$\,kHz audio with hop size $512$ and window of $2048$ samples, hence, yielding 47 frames per second.
Mean and standard deviations are calculated offline for the whole dataset and used as fixed normalization factors for the input.
The normalized mel spectrogram passes through a 1$\times$1 convolutional layer with 128 channels to extract features for the encoder.
The encoder is a sequence of multi-head attention (MHA), dropout, layer normalization, feed-forward and dropout layers, producing a latent vector to be quantized.
The network block is repeated $N=6$ times. 

The decoder architecture follows the same structure as the encoder.
Finally, a 1$\times$1 convolutional layer serves as a final projection layer to generate the decoded quantized mel spectrogram.
The sum of MSE and MAE losses ($\mathcal{L}_{prior}$) serves as reconstruction loss for the input mel spectrogram. 

For quantization we use a learned residual VQ based on VQ-VAE~\cite{vq_vae}, with projections to small dimensional spaces similar to~\cite{dac}.
FlowMAC uses a codebook size of 256 and 8 quantizer stages and a downsampling dimension 16 for the 128-dimensional latent.
Using 8 bits per level with 47 frames per second results in a rounded total of 3\,kbps.

\subsection{CFM Module}
The CFM architecture follows~\cite{matcha} and uses a U-Net with residual 1D convolutional blocks and transformer blocks with snakebeta activations~\cite{lee2023bigvgan}.
Finally, the output of the U-Net passes through a 1D Block consisting of a 1D convolution, group normalization and a Mish activation~\cite{misra2020mishselfregularizednonmonotonic}, after which a 1$\times$1 convolutional layer creates the final output.
The corresponding time-step embeddings use a RoPE-Embedding as in~\cite{grad_tts}.

The CFM decoder is conditioned on the decoded quantized mel spectrogram via concatenation to the input Gaussian noise to estimate the corresponding vector field.
The optimization criteria $\mathcal{L}_{\text{CFM}}$ is defined in Section~\ref{sec:fm}. 

Overall, the training objective for the whole system is then

\begin{equation}
	\mathcal{L} = \lambda_p \mathcal{L}_{prior} + \lambda_v\mathcal{L}_{q} + \mathcal{L}_{\text{CFM}},
\end{equation}

\noindent where $\lambda_p=0.01$ and $\lambda_v=0.25$ denote weighting factors for the prior and VQ-VAE loss ($\mathcal{L}_{q}$). 
To improve the CFM training, we sample the timestep $t$ according to a logit normal distribution~\cite{esser2024scalingrectifiedflowtransformers} for each mini-batch.
In addition, we train our model with a classifier-free guidance (CFG) technique~\cite{ho2021classifierfree}, where the decoded mel spectrogram condition is set to zero with a probability of $p_g=0.2$, which improves signal quality.

\subsection{Mel-to-Audio Module}
As mel-to-audio module, we re-train a smaller version of BigVGAN~\cite{lee2023bigvgan} on our data: We adapt the mel spectrogram calculation to fit the setting described in Section~\ref{subsec:enc-dec}.
Then, we decrease the decoder initial channels to 1024 and use an additional upsampling layer.
This yields a smaller architecture than the original BigVGAN. 

Notice that the dependence of our system on this mel-to-audio module for the final audio synthesis leads to a highest achievable quality dictated by BigVGAN's performance.
This is saturated by our mel spectrogram codec and our subjective evaluations confirm this phenomenon.

\subsection{FlowMAC inference}

Thanks to the residual vector quantizer we achieve bit rate scalability via dropping out codebook levels at inference time.
Moreover, the iterative nature of the Euler method used for inference enables some freedom on the number of function evaluations (NFE) for the CFM decoder.
FlowMAC works at 1.5 and 3\,kbps, uses 32 steps for the ODE solver and factor 1 for the CFG, hence, leading to a total of 64 NFE. 

Early experimentation showed that the quality of the mel coder subsystem quickly saturates.
To test this, we introduce FlowMAC-CQ, a separately trained model at 6\,kbps.
For this we use the same hyperparameters and NFE as for FlowMAC.
Finally, we test the quality-complexity trade-off via using a single step for the Euler method and no CFG, hence, obtaining FlowMAC-LC (Low Complexity) and using 1 NFE.

Informal listening showed that using more than 64 NFE did not bring significant improvement in quality.
Careful attention needs to be placed on the choice of the CFG factor: values smaller than 0.2 usually lead to noisy signals (except for the single-step Euler method) and values bigger that 2 overestimate the energy and introduce unwanted artifacts.

\section{Evaluation}
\subsection{Experimental setup}
We train both FlowMAC and BigVGAN on a combination of the full LibriTTS~\cite{libri_tts} clean and dev train subsets as in~\cite{lee2023bigvgan} and an internal music database consisting of $640$ hours of high-quality music of various genres.
The sampling rate for all datapoints was $24$\,kHz.

BigVGAN was trained following the official implementation~\cite{bigvgancode} for 1M iterations on a single A100 GPU. 
FlowMAC was trained with the Adam optimizer with learning rate $10^{-4}$, a segment length of 2\,s and batch size of 128 for 700k iterations on a single RTX3080.

\begin{table}[]
	\centering
\caption{Complexity measurements. Numbers for FlowMAC and FlowMAC-LC include BigVGAN. RTF is the ratio between the inference and the input duration measured on a notebook with Intel Core i7-10850H CPU @ 2.70GHz.}
	\begin{tabular}{@{}lcc@{}}
		\toprule
		Model & Nr. of Params & RTF  \\ \midrule
		DAC & 75\,M &  \phantom{0}1.24 \\
		MBD & 15\,M & 50.55 \\	
		FlowMAC & 80\,M & \phantom{0}3.38 \\
		FlowMAC-LC & 80\,M & \phantom{0}0.78 \\ \bottomrule
	\end{tabular}
	\label{tab:cx}
\end{table}

\subsection{Subjective Evaluation}
To evaluate the proposed system, we perform a P.808 DCR~\cite{p808} listening test with naive listeners and a MUSHRA~\cite{mushra2015} listening test with expert listeners. 
To this end, we design a test set of 12 items carefully selected to represent typical challenging signals for audio codecs.
The test set includes 4 clean and noisy speech samples (male, female, child and speech over music, including fast and emotional speech with varying prosody), 5 music items of various genres (including rock, pop, classical and electronic), and 3 out-of-distribution items (castanets, harpsichord and glockenspiel).

For the P.808 DCR test we compare FlowMAC to state-of-the-art DNN-based audio codecs and a well-known legacy codec.
We select the GAN-based audio codecs DAC~\cite{dac} and EnCodec~\cite{defossez2023encodec}, and the DDPM-baseline MBD~\cite{NEURIPS2023_MDB}.
We use the official implementations and pre-trained weights for all those models \cite{copet2023simple,daccode}.
We recognize that the training sets vary strongly between the conditions.
Still, we consider it useful to compare our system with well-established and robust codecs.

As a measure of the highest achievable quality with FlowMAC we include the copy-synthesis of the signals via BigVGAN.
As a benchmark legacy-condition we use an internal implementation of the MPEG-D USAC Standard~\cite{neuendorf2013the}.
This works on full-band audio, but we downsample the decoded signal to 24\,kHz to more closely measure the differences in the codecs at this sample rate.
This puts USAC at a disadvantage and may result in lower scores for it.
As a lower anchor a low-pass filter with cutoff frequency of 3.5\,kHz was used.

The P.808 DCR offers a good overall idea of the average quality of the different conditions in the test.
The MUSHRA test provides finer comparisons between a subset of the most promising conditions.
Therefore, BigVGAN, FlowMAC at 3\,kbps, DAC at 6\,kbps, MBD at 6\,kbps and USAC at 8\,kbps are selected for the MUSHRA test.

\subsection{Complexity}
We measure the complexity of the DNN-based codec systems included in the MUSHRA listening test and FlowMAC-LC in terms of numbers of parameters and real-time factor (RTF).
Table~\ref{tab:cx} summarizes the results.
The only condition able to generate the audio faster than real time is FlowMAC-LC.
We do not report the complexity figures for USAC, but we notice it is significantly faster than the DNN-based codecs.
The implementation of the DNN codecs are not optimized for synthesis speed.
We notice that none of the codecs is able to operate at low algorithmic delay, hence faster than real time generation would not enable the application of these codecs in telecommunications.

\section{Results and Discussion}

\begin{figure}[t]
  \centering
  \includegraphics[width=\linewidth]{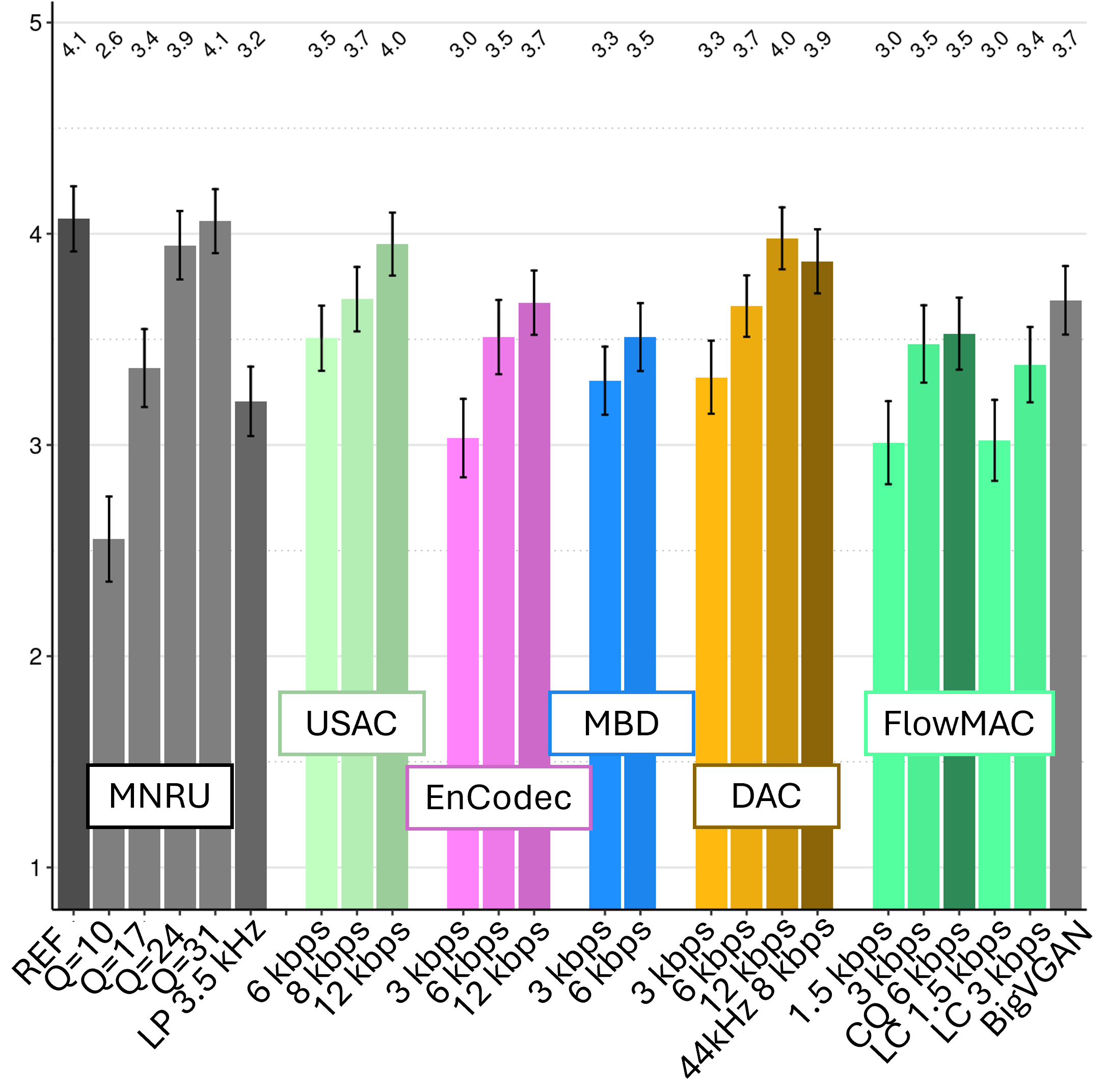}
  \caption{Results for P.808 DCR test with 46 listener and 95\% CI.}
  \label{fig:p808}
\end{figure}

\begin{figure}[t]
  \centering
  \includegraphics[width=\linewidth]{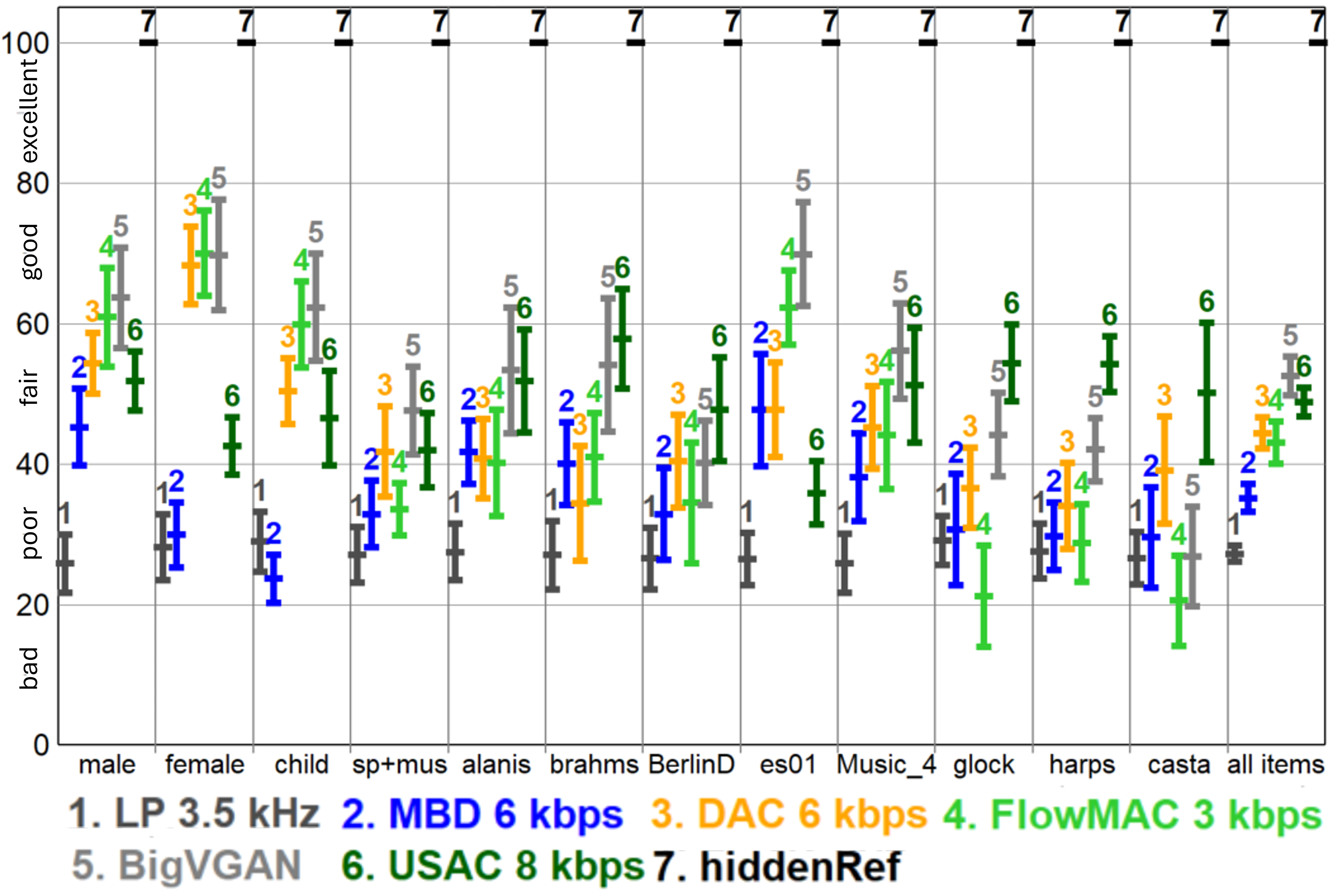}
  \caption{Results for MUSHRA test with 14 listeners and 95\% CI.}
  \label{fig:mushra}
\end{figure}

Figure~\ref{fig:p808} illustrates the results of the P.808 DCR listening test with 46 listeners.
The results from the naive listeners confirm that both FlowMAC and FlowMAC-LC are the best models at 3\,kbps, being on average on par with EnCodec and MBD at 6\,kbps.
FlowMAC at 1.5\,kbps shows a significant quality drop, while no significant quality improvement is achieved by the 6\,kbps version FlowMAC-CQ.
As expected, the copy-synthesis with BigVGAN offers the highest achievable quality for our system.
FlowMAC-LC's average rating are lower than the high-complexity version. Still, the test confirms that it is a competitive baseline.
For naive listeners the higher frequency resolution of DAC 44.1\,kHz at 8\,kbps does not offer an advantage over the 24\,kHz model.

Overall we notice that all DNN conditions achieve comparable quality with the legacy USAC condition at similar bit rates, the only exception being FlowMAC at 3 kbps.

The results of the MUSHRA test with 14 listeners are illustrated in Figure~\ref{fig:mushra}.
While USAC 8\,kbps has a quality advantage on average over the other codecs here, this test demonstrates that FlowMAC at 3\,kbps performs similar to DAC 6\,kbps and both conditions outperform MBD 6\,kbps.
We notice that the performance of the DNN-based codecs highly varies for different items in the test set.
In particular, FlowMAC performs poorly on the out-of-distribution test items, while its performance is on average comparable with DAC for speech and music.
The copy-synthesis from BigVGAN performs best average and offers a measurement of the highest quality achievable with FlowMAC.
We notice that these results more clearly highlight fine difference between the codecs, but are overall in accordance with our P.808 test results.\footnote{Check our demo samples at: https://fhgspco.github.io/flowmac}

\section{Conclusions}
This work proposed FlowMAC, a low bit rate neural audio codec for high-quality coding at 3\,kbps.
We present a novel approach for training and synthesis for a neural audio codec based on CFM.
Our subjective evaluations demonstrate that FlowMAC outperforms strong baselines while offering manageable complexity.

\section*{Acknowledgment}
We want to thank Dr. Andreas Brendel for reviewing the manuscript.
We thankfully acknowledge the scientific support and HPC resources provided by the Erlangen National High Performance Computing Center (NHR@FAU).

\bibliographystyle{IEEEtran}
\bibliography{refs}

\end{document}